# Reconfigurable nano-scale spin-wave directional coupler

Qi Wang[1], Philipp Pirro[1], Roman Verba[2], Andrei Slavin[3], Burkard Hillebrands[1], Andrii V. Chumak[1]

1. *Fachbereich Physik and Landesforschungszentrum OPTIMAS, Technische Universität Kaiserslautern, Kaiserslautern 67663, Germany*

2. *Institute of Magnetism, Kyiv 03680, Ukraine*

3. *Department of Physics, Oakland University, Rochester, MI 48309, USA*

**Abstract**

A spin-wave (SW) directional coupler comprised of two laterally parallel nano-scale dipolarly-coupled SW waveguides is proposed and studied using micromagnetic simulations and analytical theory. The energy of a SW excited in one of the waveguides in the course of propagation is periodically transferred to the other waveguide and back, and the spatial half-period of this transfer is defined as the coupling length. The coupling length is determined by the dipolar coupling between the waveguides, and the fraction of the SW energy transferred to the other waveguide at the device output can be varied with the SW frequency, bias magnetic field, and relative orientation of the waveguide's static magnetizations. The proposed design of a directional coupler can be used in digital computing-oriented magnonics as a connector (multiplexer) of magnonic conduits without a direct contact, or in the analog microwave signal processing as a reconfigurable nano-scale power divider and/or frequency separator.

**Introduction**

The research field of magnonics is attracting growing attention due to its potential for application in the next generation of information processing devices, in which information will be carried by spin waves (SW) (or magnons) instead of electrons (see reviews[1-5]). The spin waves are orders of magnitude shorter compared to the electromagnetic waves of the same frequency, and, therefore, the use of spin waves allows one to design much smaller (sometimes nano-sized) devices for both analog and digital data processing[1-14]. Recently, several novel concepts of magnonic logic and signal processing have been proposed[2,3,5, 15-24], but one of the unsolved problems of the magnonic technology is an effective and controllable connection of separate magnonic signal processing devices into a functioning magnonic circuit[5]. Unfortunately, a simple X-type crossing[20,25] has a significant drawback, since it acts as a SW re-emitter into all four connected SW channels. Thus, an alternative solution for a SW device connector is necessary.

We propose to use dipolar interaction between magnetised or self-biased, laterally parallel SW waveguides to realize a controlled connection between magnonic conduits. Originally, such a SW coupling has been studied theoretically in a "sandwich-like" vertical structure consisting of two infinite films separated by a gap[26,27]. However, the experimental studies of such structure are rather complicated due to the lack of access to the separate layers, which is required for the excitation and detection of propagating spin waves. The configuration of a connector based on two laterally adjacent waveguides is well-studied in integrated optics, since it can be conveniently implemented in applications[28]. Recently, the SW coupling in two laterally adjacent magnetic stripes has been studied experimentally using Brillouin Light Scattering (BLS) spectroscopy[29-31]. It has been shown that the SW coupling efficiency depends on both the geometry of a magnonic waveguide and the characteristics of the interacting SW modes[29]. Also, a nonlinear coupling regime of spin waves has been investigated experimentally in macroscopic scaled waveguides[30].

However, an investigation of coupling of SWs propagating in nano-sized magnonic waveguides is still missing, as well as a general analytical theory describing the SW coupling in laterally adjacent magnetic waveguides. Thus, work to study these issues and presenting a practical design of a nano-sized SW directional coupler performing effective and controllable connection between magnonic conduits is needed.

In this paper, we study the dipolar coupling of nano-scale SW waveguides with parallel and anti-parallel orientations of static magnetization using analytical theory and micromagnetic simulations, and propose a practical design of a reconfigurable SW directional coupler based on this study.

We introduce the notion of a "coupling length" $L$, defining it as a distance over which a SW propagating in one waveguide transfers all its energy to a second laterally adjacent waveguide, and study the dependence of the coupling length $L$ on the SW wavenumber (frequency), width and thickness of the waveguides, the size of the gap between them, and, finally, on the relative orientation of the static magnetization of the waveguides. We also present a practical design of a SW directional coupler containing laterally adjacent sections of nano-sized SW waveguides as a main functional element, and show their connections to the input and output waveguides used for the injection of a SW signal into the coupled waveguides with minimal reflections and optimal detection of the processed SW signal. We demonstrate by direct micromagnetic simulations that the proposed SW directional coupler can have different functionalities (connector, power divider, frequency separator, multiplexer, etc.) controlled by the external parameters, such as the frequency (wavenumber) of the propagating SW bias magnetic field, relative orientation of the static magnetizations in the coupled parallel waveguide segments, and the geometric sizes of the device. It is also demonstrated that the working parameters and the functionality of the proposed device can be dynamically reconfigured by application of a short (tens on nanosecond) pulse of an external bias magnetic field.

**General idea**

In the case, when two identical magnetic strip-line SW waveguides are placed sufficiently close to one another (see Fig. 1a), the dipolar coupling between the waveguides leads to a splitting of the lowest width SW mode of a single waveguide to the symmetric ("acoustic") and anti-symmetric ("optic") collective modes of the coupled waveguides (see Figs. 1b and 1c). Thus, in a system of two dipolarly coupled waveguides, at each frequency two SW modes (symmetric and anti-symmetric, whose profiles are shown in Fig. 1c) with different wave numbers $k_s$ and $k_{as}$ ($\Delta k = |k_s - k_{as}|$) will be excited simultaneously. The interference between these two propagating waveguide modes will lead to the periodic transfer of energy from one waveguide to the other (see Fig. 1a), so the energy of a SW excited in one of the waveguides will be transferred to the other waveguide after propagation through a certain distance that will be defined as a coupling length $L$ [27]:

$$L = \pi / \Delta k_x. \tag{1}$$

Please, note that in our case spins at the lateral boundaries are pinned only partially [40, 41] and therefore, we introduce the "effective width" of the waveguides $w_{eff}$ that can be larger than the nominal waveguide width $w$ - see Fig. 1c. This issue is discussed in more details below.

The case interesting for applications is the one when a SW is originally excited in only one SW waveguide. The output powers for both waveguides can be expressed as [27]: $P_{1\,out}=P_{in}\cos^2(\pi L_W/(2L))$ for the first waveguide, and $P_{2\,out}=P_{in}\sin^2(\pi L_W/(2L))$ for the second one, where $L_W$ is the length of the coupled waveguides as shown in Fig. 1a, and $P_{in}$ is the input SW power in the first waveguide. The dependence of the normalized output power of the first waveguide $P_{1\,out}/(P_{1\,out} + P_{2\,out})$ can be expressed as:

$$P_{1\,out} / (P_{1\,out} + P_{2\,out}) = \cos^2(\pi L_W / (2L)). \tag{2}$$

This dependence is shown in Fig. 1d as a function of the coupling length $L$ for the case when damping is ignored.

Thus, the interplay between the length of the coupled waveguides $L_W$ and the coupling length $L$, which is strongly dependent on several external and internal parameters of the system, allows one to define the ratio between the SW powers at the outputs of two coupled waveguides, and, thus, define the functionality of the investigated directional coupler. In particular, we demonstrate below that the functionality of the directional coupler can be changed by varying the geometrical sizes of the waveguides and spacing in between them, by changing the SW frequency (wavenumber) or/and the applied bias magnetic field, and changing the relative orientations of static magnetizations of the interacting SW waveguides.

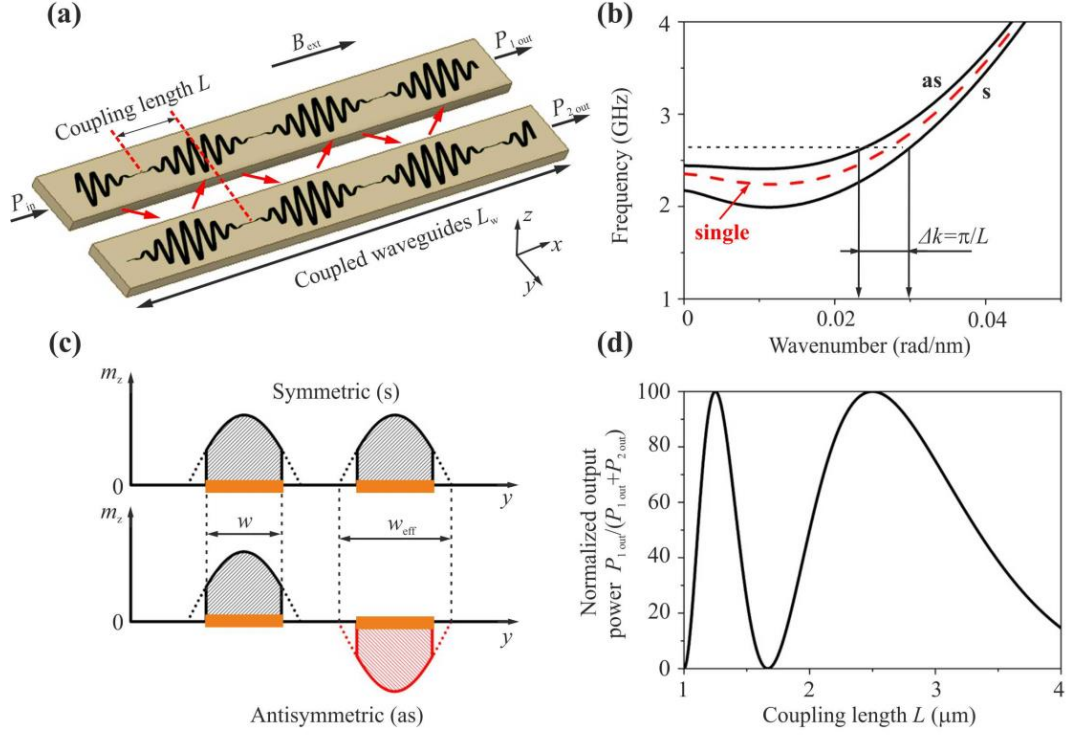

Figure 1. **The operational principle of a directional coupler based on two dipolarly coupled SW waveguides**: (a) Sketch of two dipolarly coupled SW waveguides. Solid black lines illustrate the periodic energy exchange between the two interacting SW waveguides with a spatial periodicity of 2$L$; (b) The red dashed line shows the dispersion characteristic of the lowest SW width mode in an isolated single SW waveguide. Solid black lines show the dispersion curves of the "symmetric" (s) and "anti-symmetric" (as) lowest collective SW modes of a pair of dipolarly coupled SW waveguides; (c) Spatial profiles of the "symmetric" (s) and "anti-symmetric" (as) collective SW modes partially pinned at the waveguide lateral boundaries ($w_{\text{eff}} > w$, see main text); (d) Normalized output power of the first waveguide of the directional coupler as a function of the coupling length $L$ for a fixed length of the coupled waveguides $L_w = 5$ μm and vanishing damping in the system.

**An analytical theory of coupled waveguides**

To describe analytically the energy transfer in the system of two coupled waveguides we, first, need to find the dispersion relation and the spatial profiles of the SW eigenmodes in them. When two waveguides are coupled, the SW branch existing in an isolated waveguide is split into two collective SW modes. In the simplest case of identical waveguides these collective modes are symmetric and anti-symmetric (see Fig. 1c).

To obtain the dispersion and the spatial structure of the collective modes, we use the general formalism of collective SW excitations, developed in Refs. 34, 35 and adapted to the case of two coupled magnetic waveguides. We consider two identical SW waveguides and SW modes propagating along these waveguides in the $x$ direction. The SW frequency $\omega_k$ and vector amplitudes (dynamic magnetization components) $\mathbf{m}_{k_x,p}$ can be determined from the following equation:

$$-i\omega_{k_x} \mathbf{m}_{k_x,p} = \boldsymbol{\mu} \times \sum_q \hat{\boldsymbol{\Omega}}_{k_x,pq} \cdot \mathbf{m}_{k_x,q} \tag{3}$$

where $\boldsymbol{\mu} = e_x$ is the unit vector along the direction of the static magnetization, indices $p, q = 1, 2$ enumerate waveguides, and the tensor $\hat{\boldsymbol{\Omega}}_{k_x,pq}$ has the form:

$$\hat{\boldsymbol{\Omega}}_{k_x,pq} = \gamma \left( B + \mu_0 M_s \lambda^2 \left( k_x^2 + \kappa^2 \right) \right) \delta_{pq} \hat{\mathbf{I}} + \omega_M \hat{\mathbf{F}}_{k_x}(d_{pq}), \tag{4}$$

where $B$ is the static internal magnetic field, which in our case is considered to be equal to the external magnetic field due to the negligible demagnetization along the $x$ direction, $d_{pq}$ is the distance between the centres of the two waveguides (i.e. $d_{pp} = 0$, $d_{12} = -d_{21} = w+\delta$), $\lambda = \sqrt{2A/\mu_0 M_s^2}$ is the exchange length, $\delta_{pq}$ is the Kronecker delta, and $\hat{\mathbf{I}}$ is the identity matrix. Please, note that due to the effective dipolar boundary conditions [40, 41] at the lateral boundaries of the waveguides the width profiles of the collective SW modes will be, in general, partially pinned, resulting in a non-uniform width profile of the fundamental SW mode of the waveguide $m_{y,z}(y) \sim \cos(\kappa y) = \cos(\pi y / w_{eff})$. This non-uniformity is taken into account by the effective wave number $\kappa = \pi / w_{eff}$, where $w_{eff}$ is the effective width of the waveguide (see illustration in Fig. 1c). In general, the "effective width" of the waveguides $w_{eff}$ can be substantially larger than the nominal waveguide width $w$ when the effective pinning decreases (see Fig. 1c). The magneto-dipolar interaction is described by the tensor $\hat{\mathbf{F}}_{k_x}$:

$$\hat{\mathbf{F}}_{k_x}(d) = \frac{1}{2\pi} \int \hat{\mathbf{N}}_k e^{ik_y d} dk_y \tag{5}$$

$$\hat{\mathbf{N}}_k = \frac{|\sigma_k|^2}{w} \begin{pmatrix} \frac{k_x^2}{k^2} f(kt) & \frac{k_x k_y}{k^2} f(kt) & 0 \\ \frac{k_x k_y}{k^2} f(kt) & \frac{k_y^2}{k^2} f(kt) & 0 \\ 0 & 0 & 1 - f(kt) \end{pmatrix} \tag{6}$$

where $f(kt) = 1 - (1 - \exp(-kt))/(kt)$, $k = \sqrt{k_x^2 + k_y^2}$, $t$ is the waveguide thickness, and $\sigma_k$ is the Fourier-transform of the SW profile across the width of the waveguide. In the case of almost uniform SW profile $m_{y,z}(y) = $ const, which is realized if the waveguide width is close to or smaller than the material exchange length $\lambda$, or if the effective boundary conditions are free (i.e. $w_{eff} \to \infty$), the Fourier-transform is described by the function $\sigma_k = w \operatorname{sinc}(k_y w/2)$. For any other spatially non-uniform mode with the profile $m_{y,z}(y) \sim \cos(\kappa y)$ the Fourier-transform $\sigma_k$ can be evaluated as:

$$\sigma_k = 2\sqrt{\frac{2}{1 - \operatorname{sinc}(\kappa w)}} \left[ \frac{k_y \cos(\kappa w/2) \sin(k_y w/2) - \kappa \cos(k_y w/2) \sin(\kappa w/2)}{k_y^2 - \kappa^2} \right] \tag{7}$$

Noting that the tensor $\hat{\mathbf{F}}_{k_x}(d)$ is diagonal and real (as long as the static magnetization is directed along one of the symmetry axes of the waveguide), we can obtain simple explicit expressions for the SW dispersion relations of the waveguide SW modes. The dispersion relation for the SW mode in an isolated waveguide is:

$$f_0(k_x) = \frac{1}{2\pi}\sqrt{\Omega^{yy}\Omega^{zz}} = \frac{1}{2\pi}\sqrt{\left(\omega_H + \omega_M\left(\lambda^2 K^2 + F_{k_x}^{yy}(0)\right)\right)\left(\omega_H + \omega_M\left(\lambda^2 K^2 + F_{k_x}^{zz}(0)\right)\right)} \quad (8)$$

where $\Omega^{ii} = \omega_H + \omega_M\left(\lambda^2 K^2 + F_{k_x}^{ii}(0)\right)$, $i$ = y, z, $\omega_H = \gamma B$, $\omega_M = \gamma\mu_0 M_s$, and $K = \sqrt{k_x^2 + \kappa^2}$. Noting that the magneto-dipolar interaction *between* the waveguides is small compared to the dipolar self-interaction inside the waveguide, the dispersion relations of the two collective modes (symmetric and antisymmetric) in a pair of coupled waveguides can be obtained as:

$$f_{1,2}(k_x) \approx f_0(k_x) \pm \Delta f / 2 \quad (9)$$

where the frequency separation between the symmetric and antisymmetric collective modes is given by:

$$\Delta f = \frac{\omega_M}{2\pi}\frac{\Omega^{yy} F_{k_x}^{yy}(d) + \Omega^{zz} F_{k_x}^{zz}(d)}{\omega_0(k_x)} \quad (10)$$

**Results**

*Model and simulations*

The structure of a directional coupler is schematically shown in Fig. 2a. Two parallel waveguides (length 100 μm, width $w$ ranging between 100 and 300 nm and thickness $t$ in the range from 10 to 50 nm) are placed laterally parallel with a gap $\delta$ (ranging between 10 and 100 nm). The numerical modelling of this structure is performed using the MuMax3[36] micromagnetic package with the following parameters of an Yttrium Iron Garnet (YIG) nanometer-thick film[9,37]: saturation magnetization $M_s = 1.4\times10^5$ A/m, exchange constant $A = 3.5$ pJ/m and Gilbert damping $\alpha = 2\times10^{-4}$. In our numerical simulations the Gilbert damping at the ends of the waveguides exponentially increases to 0.5 to eliminate SW reflection at the waveguide ends. Due to the ultra-low Gilbert damping of YIG the SW propagation distances in YIG waveguides reach up to dozens of micrometers[9,13,38,39], which opens up the possibility for the realization of complex integrated SW circuits. A small external magnetic field $B_{ext} = 10$ mT is applied along the long axis of the waveguides ($x$ direction in Fig. 2). In order to excite a propagating SW in one of the waveguides, a sinusoidal magnetic field $b_y = b_0\sin(2\pi ft)$ is applied at the centre of the second waveguide over the area of 20 nm in length (see yellow area shown in Fig. 2a), with the oscillation amplitude $b_0 = 1$ mT and a varying microwave frequency $f$.

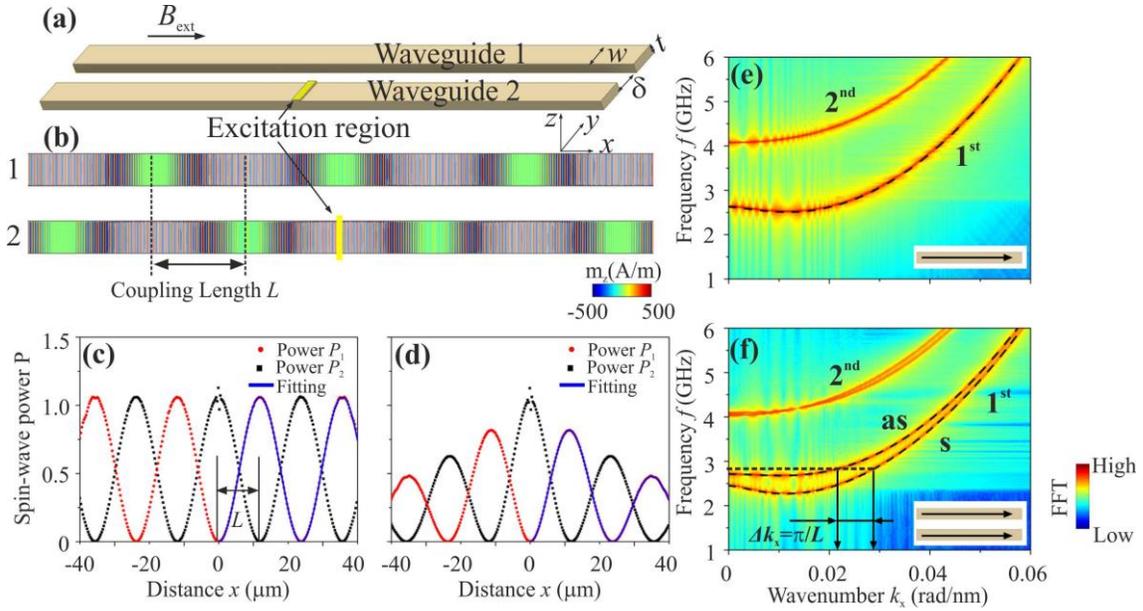

Figure 2. **Energy transfer in dipolarly-coupled waveguides.** A schematic view of the dipolarly-coupled SW waveguides (a) and a snapshot of the SW profile excited in them (b). The excitation antenna (shown in yellow) is located in the centre of one of the waveguides. The variation of the SW power in both waveguides as a function of propagation distance $x$ is shown for the case of zero SW damping (c) and for the typical damping in an YIG film (d). The dispersion characteristics of the two lowest width spin wave modes in a single isolated waveguide are shown in the frame (e), while the similar dispersion curves for a pair of coupled waveguides are shown in the frame (f). A color map represents the results of our numerical simulations, while the dashed lines represent the analytical theory. The panel (f) shows the splitting of the SW dispersion curves into anti-symmetric (as) (top) and symmetric (s) (bottom) branches due to the dipolar coupling between the waveguides.

*Splitting of SW dispersion curves due to the dipolar coupling between waveguides*

Figure 2b shows a snapshot of the dynamic magnetization profiles in the coupled waveguides with the following parameters: width $w = 100$ nm, thickness $t = 50$ nm, gap $\delta = 100$ nm. The frequency of the excited SW in this case was 2.96 GHz, corresponding to a wave number $k_x = 0.02872$ rad/nm. It can be seen, that the energy of the SW, excited in waveguide #2, is transferred completely to the waveguide #1 after the propagation over a coupling length $L$ – see Fig. 2b. Figures 2c and 2d show the variations of the SW power in the two coupled waveguides as a function of the propagation distance $x$ for the case of zero damping (frame (c)) and for the case of a typical damping of an YIG film ($\alpha = 2\times10^{-4}$) (frame (d)). The coupling length of $L = 12$ μm can be extracted from Figs. 2c and 2d. Please note that the SW wavelength in our studies is of the order of 100 nm. The simulation performed for the damping-free waveguides clearly shows the lossless oscillations of the SW energy between the waveguides. The blue line is a fit to the data using the dependence $P_1 = P_{in}\sin^2(x\pi/(2L))$, where $P_{in}$ is the power launched into the waveguide #2, and $P_1$ is the power in the waveguide #1. In the case where realistic SW damping is taken into account, the SW power gradually decreases, as expected. The results can be fitted by a similar model which includes the damping term $P_1=P_{in}\sin^2(x\pi/(2L))\exp(-|2x|/x_{freepath})$, where $x_{freepath}$ = 88 μm is the exponential decay length, which is in good agreement with the analytical theory.

The frame 2e shows the dispersion characteristics of the two lowest width modes of a single isolated waveguide and the frame and 2f shows the similar dispersion curves for a pair of waveguides coupled across a gap of $\delta$ = 10 nm. The color maps were calculated by micromagnetic simulations, while the black dashed lines for the lowest width SW mode were obtained using the analytic equations Eq. (8) and (9) for the single waveguide and the coupled waveguides, respectively.

It is seen that the dispersion curve corresponding to the lowest width mode for the case of coupled waveguides splits into two modes: antisymmetric (as) (top) and symmetric (s) (bottom) – see Fig. 1c and Fig. 2f. This splitting is caused by the dipolar interaction between the waveguides. The oscillations of the SW energy in the coupled waveguides can be interpreted as an interference of the symmetric and antisymmetric SW modes[27-29]. These two SW modes have the same frequency, but different wavenumbers, and, therefore, different phase velocities. Thus, these modes accumulate a phase difference during their propagation in the waveguides. When the accumulated phase difference is equal to π, the superposition of the two modes results in a destructive interference in one of the waveguides, and in a constructive interference in the other one. The SW energy is completely transferred from one waveguide to the other after propagation for a coupling length $L$. Analogously, the energy is transferred back to the first waveguide after the further propagation for the same distance $L$ and the process is periodic (see the Animation 1 in the Supplementary Information). The coupling length can be expressed as[27] $L = \pi/\Delta k_x$, where $\Delta k_x$ is the difference of the wavenumbers of the symmetric and the antisymmetric modes of the same frequency, as it is shown in the Fig. 2f. This difference in the wavenumbers of the modes of the same frequency is equal to $\Delta k_x = \Delta\omega/(\partial\omega_0/\partial k_x) = \Delta\omega/v_g$. Using this last expression, the coupling length can be rewritten as:

$$L = v_g / 2\Delta f, \qquad (11)$$

where $v_g$ is the group velocity of the signal spin wave in a single isolated waveguide. Note that the periodic exchange of the full SW energy between the coupled waveguides is possible only if the waveguides are identical, have the parallel or anti-parallel static magnetization and experience the same internal magnetic field. In other cases, the SW energy is also transferred on a distance $L$, but the transfer is incomplete.

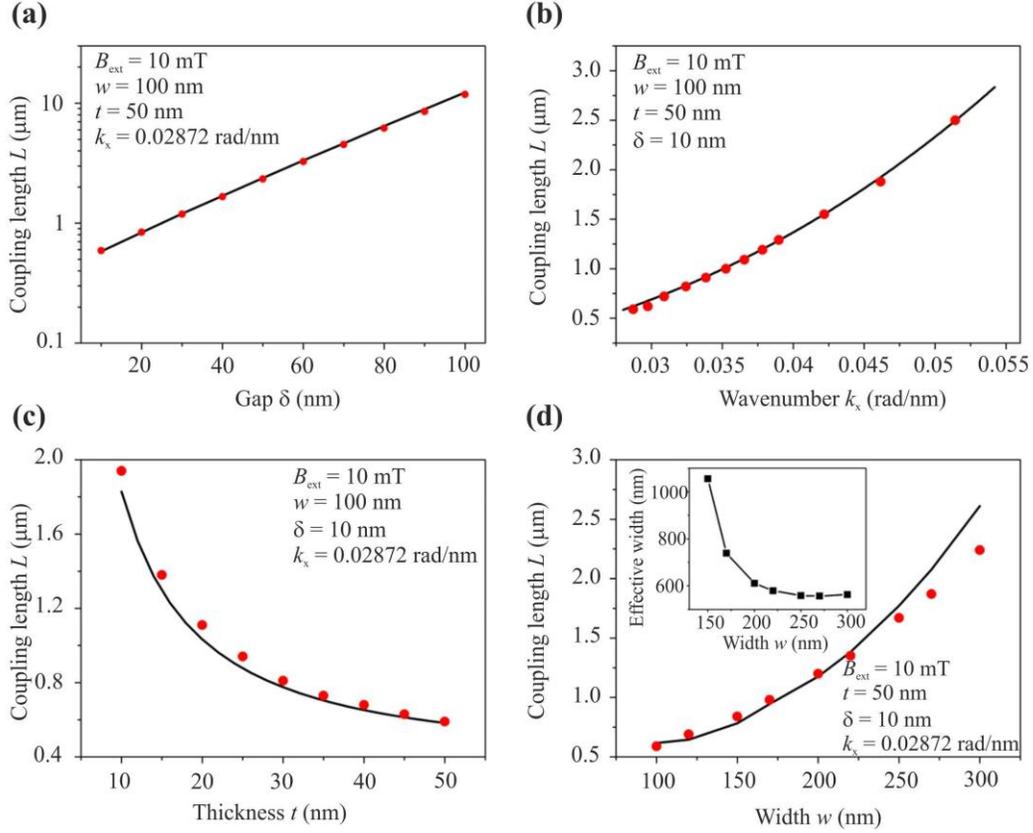

Figure 3. **Coupling length**. The coupling length $L$, i.e. the SW propagation length over which the SW energy is transferred from one waveguide to the other, is shown as a function of (a) the separation $\delta$ between the waveguides, (b) the longitudinal wavenumber $k_x$, (c) the thickness $t$ of the waveguides, and (d) the width $w$ of the waveguides. These results were obtained by means of numerical simulations (symbols) and analytical calculations (lines). The parameters for each particular case are shown directly inside the panels. The inset in panel (d) shows the dependence of the effective width $w_{eff}$ of the waveguide on the nominal width $w$ (the line is a guide for the eye). The decrease in the waveguide width results in the reduction of the effective dipolar pinning at the lateral edges of the waveguides, and in the corresponding increase of the effective waveguide width.

*Coupling length*

Figure 3 shows the dependence of the coupling length $L$ on the geometrical parameters of the waveguides: the gap size $\delta$ (Fig. 3a), the thickness $t$ (Fig. 3c), the width $w$ (Fig. 3d) of waveguides, and on the SW wavenumber $k$ (Fig. 3b). In general, it can be clearly seen that the increase of the dipolar coupling between the waveguides results in a decrease of $L$. The coupling between the waveguides is increased by a decrease of the gap $\delta$ between the waveguides, by a decrease in their width $w$, by an increase in their thickness $t$, and, finally, by a decrease in the SW wavenumber $k_x$. The results obtained from the micromagnetic simulations (circles points) and from the analytical theory (lines) are consistent. In the waveguides with a width $w > 150$ nm the SW profile across the width become essentially non-uniform due to dipolar pinning[32,33], and the micromagnetically calculated SW profiles (in an isolated waveguide) are used for the calculation of the geometric form-factor $\sigma_k$.

In the course of our simulations a single-frequency SW was excited in one of the waveguides, and the width profile of this SW mode was extracted from the results of numerical simulation. The function $m_z(y) = Amp \cdot \cos(\pi y/w_{\text{eff}})$, where $Amp$ is the amplitude, was used to fit the numerically calculated width profile of the SW mode in order to obtain the effective waveguide width $w_{\text{eff}}$. Then, this effective width was plugged into Eq. (7) for analytic calculations. The inset in Fig. 3d shows the dependence of the effective width on the geometrical width $w$ of the waveguides. The effective width strongly increases with the decrease in $w$ which corresponds to the gradual decrease of the effective dipolar pinning at the lateral edges of the waveguides [32,33].

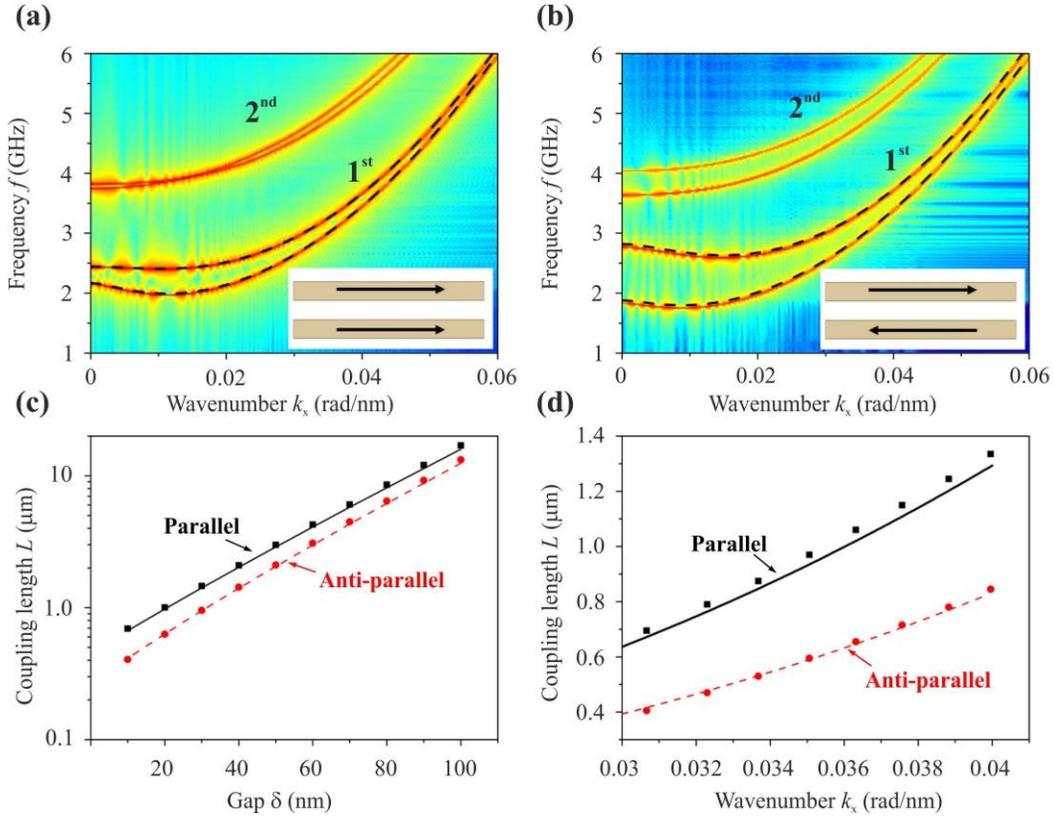

Figure 4. **Dependence on the static magnetization configuration**. Dispersion characteristics of the lowest SW width modes in coupled waveguides for the cases of parallel (a) and anti-parallel (b) orientations of the static magnetizations of the waveguides (see insets). Dashed lines show the results of the analytic theory. The waveguide parameters are: widths $w$ = 100 nm, thicknesses $t$ = 50 nm, gap width $\delta$ = 10 nm. The coupling length as a function of the gap width $\delta$ is shown in frame (c), and as a function of the longitudinal wavenumber $k_x$ - in frame (d) for parallel (black squares and solid black lines) and anti-parallel (red circles and dashed red lines) static magnetization configurations. The symbols show the numerical results, while lines correspond to the results calculated analytically.

*Anti-parallel magnetization configuration*

In the previous sections, it was assumed that a small external bias magnetic field was applied along the $x$

direction in order to orient the static magnetization along the waveguides' long axes. However, in nano-scale waveguides the static magnetization orients itself parallel to the waveguides spontaneously without any external field due to the strong shape anisotropy of the elongated nature of a waveguide. Moreover, in the absence of an external field waveguides can exist in two stable magnetic configurations – with parallel and anti-parallel static magnetizations. The analytical calculations show that the SW dispersion and, therefore, the coupling length, depend significantly on the static magnetization configuration. In particular, for the anti-parallel configuration of the waveguide static magnetizations the splitting of the dispersion relation of the symmetric and antisymmetric collective modes is given by:

$$\Delta f = \frac{\omega_M}{2\pi} \frac{\Omega^{zz} F_{k_x}^{yy}(d) - \Omega^{yy} F_{k_x}^{zz}(d)}{\omega_0(k_x)} \quad . \tag{12}$$

This equation is substantially different from the splitting Eq. (10) that takes place in the case of parallel static magnetization of the coupled waveguides. As one can see from Figs. 4a and 4b, the frequency splitting is stronger for the anti-parallel magnetization configuration which results from the stronger interaction of the oppositely precessing dynamic magnetizations in the two coupled waveguides. Consequently, the coupling length for the anti-parallel configuration is always *smaller* than for the parallel magnetization configuration, as one can see from the dependences of the coupling length $L$ on the gap $\delta$ and on the SW wavenumber $k_x$ shown in Figs. 4c and 4d, respectively.

### *Design of a directional coupler*

The dipolarly-coupled SW modes in parallel waveguides have a large potential for applications. The functionality of a microwave signal processing device based on two laterally parallel coupled waveguides depends on the ratio between the coupling length $L$ and the length of the coupled waveguides $L_W$. Thus, if $L_W = (2n+1)L$, where n is an integer value, the entire energy will be transferred from one waveguide to the other, and the directional coupler can be used as a connector of magnonic conduits. If $L_W = (n+1/2)L$ the coupler can be used as an equal divider (3dB in each beam pass) for microwave power. Taking into account that the coupling length $L$ strongly depends on the SW wave number $k$, and, consequently, on the signal frequency, the directional coupler can be used as a frequency separator. Finally, the variation of the external bias magnetic field or/and of the direction of static magnetization in one of the coupled waveguides allows for the switching of the functionality of the directional coupler having a fixed length $L_W$ and fixed signal frequency.

One of the challenging tasks in the practical realization of a direction coupler is the design of the inputs (outputs) to (and from) the coupled SW waveguides that are needed to precisely define the $L_W/L$ ratio (note e.g. the absence of such inputs and outputs in Refs. 29-31). The design of optical directional couplers[28] cannot be simply copied in magnonics, because of the anisotropy of the SW dispersion laws and their qualitative dependence of the orientation of the static magnetization in a SW waveguide[20,40]. Moreover, the SW spectra typically have a multi-mode character which can significantly complicate the SW microwave signal processing[18]. Most of these problems are automatically solved when the sizes of the magnonic signal processing devices are scaled down to below a micrometer. In this case, the frequencies of the SW modes are well separated due to the strong exchange interaction that shifts the frequency of the higher order thickness and width modes by several GHz (see e.g. Fig. 2e). Moreover, as it was described above, the strong shape anisotropy of the elongated SW waveguides makes the waveguides static magnetization parallel to the long axis of the waveguide, and, therefore, along the direction of the SW propagation, even in the absence of external bias magnetic field (see black arrows in Fig. 5a showing the direction of static magnetization). Such quasi-isotropic condition for the SW propagation is one of the big advantages of the nano-sized magnonic conduits, when compared to the similar systems of the micro- and macro-scale.

Therefore, the nano-scale structure shown in Fig. 5a, which is analogous to the directional coupler used in integrated optics[28], is also suitable for the realization of a nano-scale SW directional coupler. The main drawback of such a design is that the corners of the structure (marked with red circles in Fig. 5a) could act as secondary SW sources, thus disrupting the operational characteristics of the device. In order to minimize these distortions, we introduced in the practical model of the proposed directional coupler a *translational shift* $d_s = 100$ nm between the beams, as it is shown in Fig. 5a. In the following studies, we fixed the width of the waveguides to $w = 100$ nm, the thickness to $t = 50$ nm, and the gap between the waveguides to $\delta = 30$ nm. The length of the coupled waveguides $L_W$ was fixed to be equal to $L_W = D - d_s = 4900$ nm.

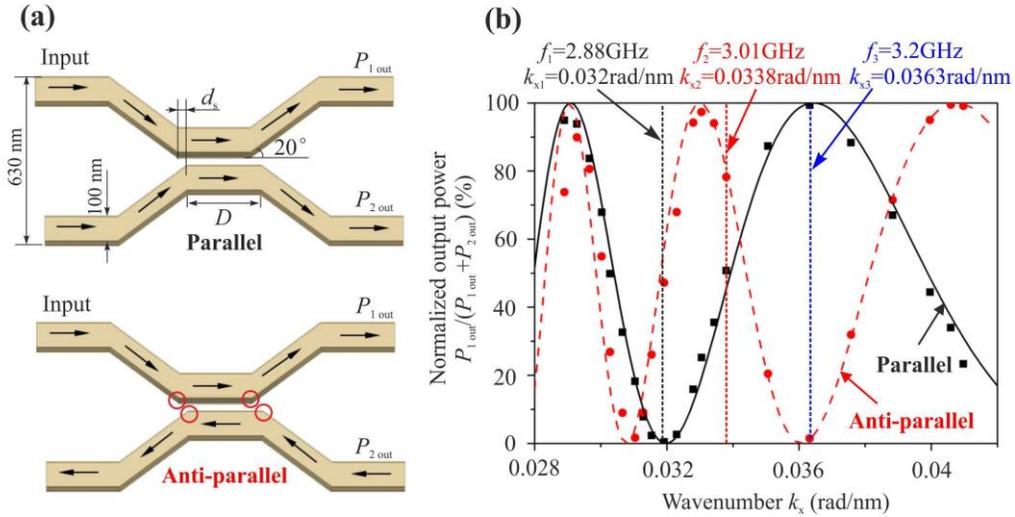

Figure 5. **Reconfigurable SW directional coupler:** (a) Schematic view of the parallel and anti-parallel magnetization configuration of the directional coupler. The widths of the waveguides are $w = 100$ nm, thickness is $t = 50$ nm, gap is $\delta = 30$ nm, the angle between the coupler waveguides is 20 degrees, the working length of the coupled waveguides is $L_W = D - d_s = 4900$ nm. The arrows show the direction of the static magnetization. The spin waves are excited in the first beam pass of the directional coupler marked as "Input"; (b) The wave number dependence of the normalized power at the output of the first beam pass $P_{1\,out}/(P_{1\,out}+P_{2\,out})$ of the directional coupler. The symbols and lines were obtained by micromagnetic simulations and the analytical theory for parallel (black squares and solid black line) and anti-parallel (red circles and dashed red line) configurations, respectively. The vertical dashed lines indicate the wavenumbers (and corresponding frequencies) which are chosen for the demonstration of different functionalities of the directional coupler in Fig. 6.

Figure 5b shows the normalized output power in the first beam pass $P_{1\,out}/(P_{1\,out}+P_{2\,out})$ as a function of the SW wavenumber $k_x$. The symbols represent the results of the micromagnetic simulations and the lines are obtained from the analytical theory for the parallel (black) and anti-parallel (red) configuration of the waveguides static magnetizations, respectively.

Please note that, in spite of the fact that the directional coupler as a whole was studied by micromagnetic simulations, while only the parallel parts (of the length $L_W$) of the coupled waveguides were considered in the framework of the analytical theory, the difference in the results is negligible. This is an indication of the high efficiency of the proposed input and output waveguide structures, which, obviously, demonstrate small reflections for the propagating SWs. It can be clearly seen in the Figure 5, that the operational characteristics of the directional coupler can be easily tuned by the variation of the SW wavenumber (i.e. by the frequency of the input microwave signal).

Moreover, this tuning is different for parallel and anti-parallel configurations of the static magnetization. In the following, we choose three different values of the SW wavenumber, which are shown in Fig. 5b by vertical dashed lines, and perform separate simulations for the case of a single-frequency input signal. The design and all the sizes of the directional coupler are kept the same for all the simulations shown in Fig. 6.

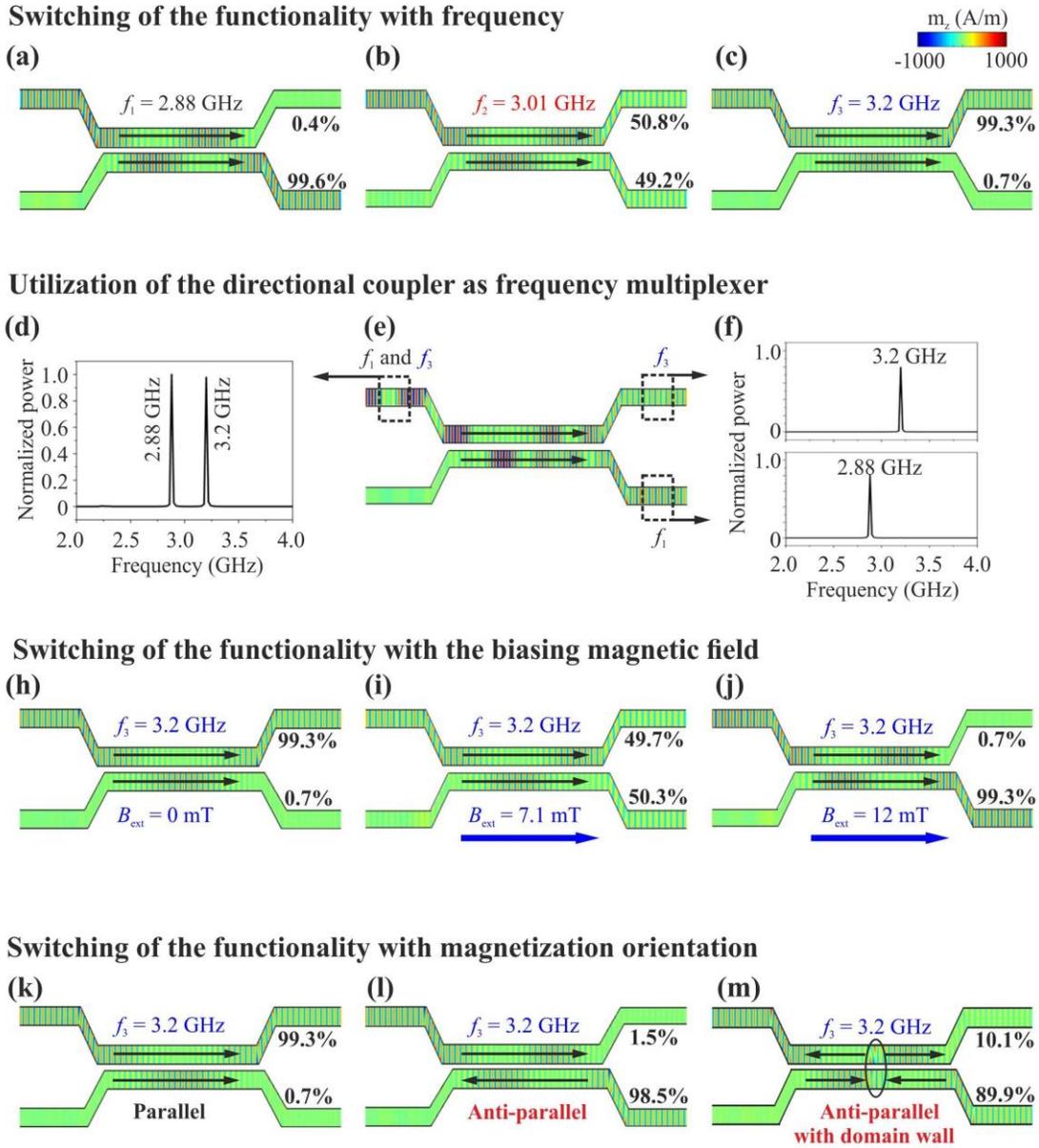

Figure 6. **Dynamically reconfigurable SW directional coupler**
*Switching of the device functionality by changing the signal frequency*: (a)-(c) Directional coupler acts as a connector of magnonic conduits, as a 3 dB power divider, or as a simple transmission line.
*Utilization of the directional coupler as frequency separator (multiplexer)* (d)-(f). The directional coupler can be used as a frequency multiplexer: SWs of two frequencies simultaneously excited in the first beam pass of the coupler will reach different output beams.
*Switching of the device functionality by changing the bias magnetic field*: (h)-(j) The ratio of the output powers in two beam passes can be changed by the variation of the bias magnetic field.
*Switching of the device functionality by changing the static magnetization orientation*: (k)-(l) Switching the relative orientation of the static magnetization in two beam passes leads to the switching of the output signal between the beam passes; (m) Even in the case when the remagnetization process leads to the formation of domain walls in the device beam passes, the main part (90%) of the output signal power is still switched from the upper to the lower beam pass.

The SW amplitude is shown by a color map. Note that the width of the waveguides is constant in all parts of the directional coupler, as is shown in Fig. 5a. The structures in Fig. 6 are compressed in the direction along the waveguide for a better illustration of the coupling effects.

*Functionalities of a directional coupler*

Figure 6 shows the color maps of the SW amplitude (represented by the variable component of the dynamic magnetization $m_z$) in a directional coupler for different input frequencies and, consequently, wavenumbers. As it is expected from the results shown in Fig. 5b, the SW of the frequency $f_1 = 2.88$ GHz is almost fully transferred to the second waveguide. After a few oscillations between the waveguides (see Fig. 6a) 99.6 percent of the output energy is exited from the second pass of the device in our simulations. The coupling length ($L = 1630$ nm in this case) satisfies the ratio $L_W = 3 \times L$. Thus, this directional coupler can be used to effectively connect two magnonic conduits. If the spin wave of the same frequency is excited in the other pass of the coupler, the SW energy will be transferred into the opposite pass in a similar way.

The situation is different for the spin wave of the frequency $f_3 = 3.2$ GHz (see Fig. 6c) that corresponds to a longer coupling length of $L = 2450$ nm. The length of the waveguides is $L_W = 2 \times L$ in this case (see also Fig. 5b), and the SW energy is transferred back to the input pass of the directional coupler. That means that this directional coupler can be used as a frequency separator (multiplexer): if SWs of different frequencies $f_1$ and $f_3$ are simultaneously excited in the same beam pass of the waveguide, the SW of frequency $f_1$ will exit from one pass of the coupler, while the signal of the frequency $f_3$ will exit from the other pass, as shown in Figs. 6d-f (see also Animation 2 in the Supplementary information). Figure 6b shows that the directional coupler can be used as a 3 dB power divider in which half of the energy is transferred to the second pass of the coupler and half of the energy stays in the first pass. A propagating SW of the frequency $f_2 = 3.01$ GHz corresponding to $L = 1960$ nm has been excited in this case in order to ensure the condition: $L_W = 2.5 \times L$. The ratio between the output energies in both beam passes of the waveguide can be easily tuned by the frequency of the input signal.

Furthermore, the ratio between the output powers can be adjusted by shifting the dispersion curves up or down using an applied external magnetic filed. A spin wave of the frequency $f_3 = 3.2$ GHz has a coupling length $L = 2450$ nm (unbiased case). In this case, the SW energy is transferred back to the same input pass of the directional coupler – see Fig. 6h. When the external bias magnetic field is increased to 7.1 mT, the coupling length decreases to 1960 nm, and the directional coupler acts as a 3 dB power divider – see Fig. 6i. The continuing increase of the bias field to 12 mT results in the coupling length further decreasing to 1630 nm, and, in such a situation, most of the SW energy is transferred to the second pass of the directional coupler, as it is shown in Fig. 6j. The energy of the propagating SW can be switched from one pass to the other during a few nanoseconds using the application of an abrupt step in the external bias field (see Animation 3 in the Supplementary information).

Finally, we study the case when a spin wave of frequency $f_3$ is excited (see Fig. 6k), but the relative orientation of the static magnetizations in the beam passes has been switched from parallel to anti-parallel (see Fig. 6l). As expected based on the results presented in Fig. 5b, the majority of the SW energy is transferred in that case from one pass of the directional coupler to the other one due to the shorter coupling length $L$ for the anti-parallel magnetization configuration ($L$ = 1640 nm in this case satisfies the condition $L_W = 3 \times L$). Thus, the proposed directional coupler turns out to fully dynamically reconfigurable, and can be used as an effective and fast switch or/and multiplexer[8].

Obviously, the operational frequency of the device can be easily varied with the length of the directional coupler $L_w$, by the geometry of the waveguides that defines the coupling length $L$, or by the external bias magnetic field. It is important to note, however, that the maximum operating power of such a frequency multiplexer is limited by different non-linear SW phenomena[17,29,40].

### *Switching of the orientation of static magnetization in a directional coupler*

Finally, we would like to discuss the practical realization of the switching of the relative orientation of the static magnetizations in the passes of the directional coupler (see Fig. 4). Taking into account a very short separation distance between the parallel passes of the coupler, it is difficult to re-magnetize the passes of the coupler independently using external magnetic fields. Instead, here we propose to use a method similar to the one used in Refs. [11,41-43] for the switching of the magnetic state in arrays of dipolarly-coupled magnetic nano-dots. Namely, we apply a short (~20 ns) magnetic field pulse in the perpendicular direction ($y$ direction), which magnetizes both waveguides in the $y$ direction (Fig. 7a).

Then, the applied field is decreased to zero within 10 ns. This evolution allows the static magnetization in the coupler passes to be spontaneously directed along the long axes of the waveguides (see the Animation 4 in the Supplementary Information), but also results in the excitation of parasitic spin waves in the structure (see the long tail in the $M_y$ characteristics). In general, the device is ready for operations after a time period exceeding at least one SW life cycle, which, in our case, is equal to 252 ns (for the 3 GHz frequency).

However, even 50 ns after the bias field has been tuned off, one can clearly see that the magnetization distribution in the directional coupler assumes the form shown in the right panels of Fig. 7a. Each pass of the waveguide contains a domain wall in the centre of the structure, and is, therefore, separated into two regions with opposite directions of the static magnetization. The magnetization orientation in both regions is anti-parallel to the magnetization orientation in the adjacent pass of the directional coupler, as desired.

In order to prove that the domain walls do not substantially disturb the operational characteristics of the device, additional simulations for a single SW waveguide with the same domain wall structure have been performed and have shown that the SW reflection coefficient due to the domain wall existence is only 3%. Figure 6m demonstrates that the magnetization configuration with the presence of the domain walls does not have much influence on the operational characteristics of the directional coupler, when compared to the ideal anti-parallel aligned magnetization configuration shown in Fig. 5l.

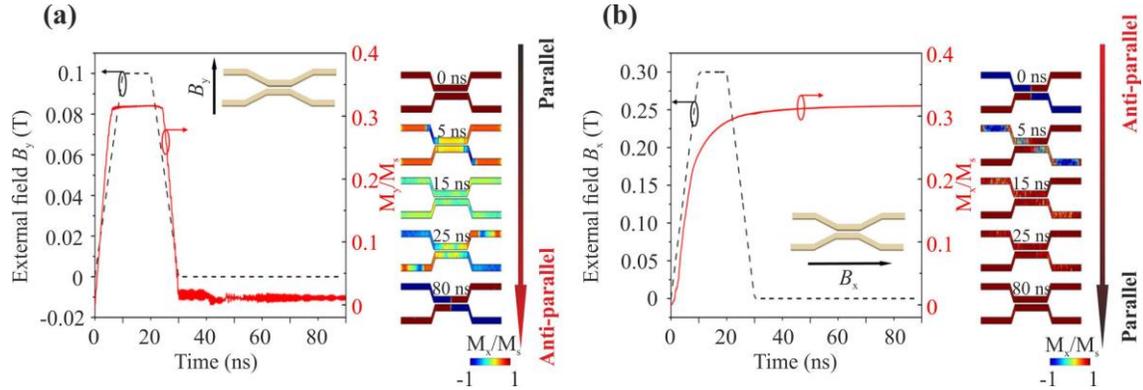

Figure 7. **Switching of the directional coupler.** (a) Application of a magnetic field pulse in the direction transverse to the coupled waveguides direction results in the switching of the magnetization from the parallel to the anti-parallel state – see color map in the right panels for $M_x$ component. Two domain walls are formed in the centres of the coupler beam passes in such anti-parallel magnetization configuration; (b) Profile of the longitudinal magnetic field pulse for the switching of the directional coupler magnetization back into the parallel state. Magnetization configuration is shown in the same way as in the Panel (a).

In order to switch the magnetization configuration of the directional coupler back to the parallel state, we applied a field of 0.3 T in the direction *x* (parallel to the waveguides) as it is shown in Fig. 7b. The magnetic field is switched on again for 30 ns with a rise and a fall time of 10 ns. It is important to note that these time intervals are important since the switching of the magnetic field for a shorter period will not necessarily result in the switching of the magnetization configuration. The magnetization orientation is shown in the same way as in the Fig. 7a. One can clearly see that approximately 50 ns after the external field was switched off, the directional coupler stays in its original parallel magnetization configuration (see the Animation 5 in the Supplementary Information). Figure 6k shows the SW amplitude in this case, and one can clearly see that the SW energy reaches the same beam pass in which it was originally excited. Thus, the proposed methodology allows for the realization of the SW switch.

We would like to mention that the simulations presented here were performed without taking into account temperature, i.e., for zero effective temperature. Additional numerical simulations, identical to those which are shown in Fig. 6a, were performed at an effective temperature of 300 K in order of explore the influence of temperature on the characteristics of the directional coupler – see Animations 6 and 7 in the Supplementary Information. Only a small difference between the operational characteristics of the device was obtained for different temperatures: the output power of the device is 99.6% at zero temperature and 86.3% at room temperature. The difference is mainly caused by a slight shift of the dispersion characteristics of SWs that results, consequently, in a change of the coupling length. We expect that the role of temperature can be decreased by an adjustment of the coupling length via a slight variation of the signal frequency. The output power of the device can be increased to 94.5% at room temperature by slightly decreasing the frequency to 2.84 GHz – see Animation 8 in the Supplementary Information.

**Conclusion**

In conclusion, the practical design of a nano-scale SW directional coupler is presented and studied by means of micromagnetic simulations. The interference between the two collective SW modes of two laterally parallel and dipolarly coupled magnetic waveguides separated by a gap provides the mechanism responsible for the operation of the device. A general analytical theory describing such coupling has been developed. The coupling length $L$, over which a SW transfers all its energy from one waveguide to the other, is studied as a function of the SW wavenumber, geometry of the coupler, and the relative orientation of the static magnetization in the waveguides. The design which allows the directional coupler to be used as a connector of magnonic conduits, as a magnonic switch and as a frequency separator (multiplexer), as well as a power divider for microwave signals have been proposed. Finally, a practical way for the switching of the functionality of the directional coupler within few tens of nanoseconds is proposed and proven by means of micromagnetic simulations. Taking into account the nanometer size of the proposed directional coupler, this device will be of interest for the processing of both digital and analogue microwave information.

**Methods**

*Extraction of the dispersion relations from the results of micromagnetic simulation*

The micromagnetic simulations were performed using the MuMax3[36] code. It employs the Dormand-Prince method[44] for the integration of the Landau-Lifshitz-Gilbert (LLG) equation:

$$\frac{d\mathbf{M}}{dt} = -|\gamma|\mathbf{M} \times \mathbf{B}_{\text{eff}} + \frac{\alpha}{M_s}\left(\mathbf{M} \times \frac{d\mathbf{M}}{dt}\right) \qquad (11)$$

where $\mathbf{M}$ is the magnetization vector, $\mathbf{B}_{\text{eff}}$ is the effective magnetic field, $\gamma$ is the gyromagnetic ratio, and $\alpha$ is the damping constant. The material parameters have been given in the main text. There are three steps involved in the calculation of the SW dispersion curve in our simulation. The external field is applied along the waveguide, and the magnetization is relaxed to a stationary state. This state is consequently used as the ground state in the following simulations. In order to excite odd and even SW width modes, a sinc field pulse is applied to a 20 nm wide area located on one side of the waveguide. The sinc field is $b_y=b_0\text{sinc}(2\pi f_c t)$, with oscillation field $b_0 = 1$ mT and a cutoff frequency $f_c = 20$ GHz. The $M_z(x,y,t)$ of each cell is collected over a period of $T = 100$ ns and recorded in $T_s = 25$ ps intervals, which allows for a frequency resolution of $\Delta f = 1/T = 0.01$ GHz while the highest resolvable frequency is $f_{\text{max}} = 1/(2T_s) = 20$ GHz. The fluctuations in $m_z(x,y,t)$ were calculated for all the cells, $m_z(x,y,t) = M_z(x,y,t) - M_z(x,y,0)$, where $M_z(x,y,0)$ corresponds to the ground state obtained from the first step.

In order to obtain the SW dispersion curves, we performed 2D Fast Fourier transformation[45,46]

$$m_z(k_x, f) = \frac{1}{N}\sum_{i=1}^{N}\left|\mathcal{F}_2\left[m_z(x, y_i, t) - m_z(x, y_i, 0)\right]\right|^2 \quad (12)$$

where $\mathcal{F}_2$ is the 2D Fast Fourier transformation (FFT) and $y_i$ is the $i$-th cell along the width of the waveguide, $N$ is the total number of the cells along the width of waveguide. In order to visualize the dispersion curve we recorded a 3D color map of $P(k_x, f) \propto m_z(k_x, f)$ in logarithmic scale versus $f$ and $k_x$ which is shown in Figs. 2e-f and Figs. 4a-b. We performed 2D FFT on the time evolution, and along the waveguide. Next, the average FFT amplitude was taken along the width of the waveguide. This method allows to obtain information about all the SW modes (even and odd) existing in the waveguide.

**Acknowledgements**

This research has been supported by the ERC Starting Grant 678309 MagnonCircuits. R. V. acknowledges support from Ministry of Education and Sciences of Ukraine (project 0115U002716). A.S. was supported by the Grants No. ECCS-1305586 and No. EFMA-1641989 from the National Science Foundation of the USA, by the contract from the US Army TARDEC, RDECOM, and by the Center for NanoFerroic Devices (CNFD) and the Nanoelectronics Research Initiative (NRI).

**Author Contributions**

Q. W. proposed the spin-wave directional coupler design, performed micromagnetic simulation, and wrote the first version of the manuscript. P. P. and A. V. C. devised and planned the project. R. V. and A. S. developed the general analytical theoretical model. A. V. C. led the project. All authors discussed the results and contributed to writing the manuscript.

**Author Information**

The authors declare no competing financial interests. Correspondence and requests for materials should be addressed to A.V.C. (chumak@physik.uni-kl.de)